\documentclass[letterpaper,twocolumn,10pt]{article}
\usepackage{usenix-2020-09}

\usepackage{tikz}
\usepackage{amsmath}

\usepackage{filecontents}

\usepackage{subcaption}

\usepackage{hyperref}

\usepackage{pgfplots}
\usepgfplotslibrary{statistics}
\pgfplotsset{width=\textwidth,compat=1.18}

\usepackage{enumitem}

\begin{document}

\date{}

\title{\Large \bf An intensive vRAN deployment with OpenAirInterface}

\author{
{\rm Romain Beurdouche}\\
EURECOM, Biot, Provence Alpes Côte d'Azur, France
\and
{\rm Raymond Knopp}\\
EURECOM, Biot, Provence Alpes Côte d'Azur, France
} 

\maketitle

\begin{abstract}
  The advent of 5G virtualized Radio Access Networks (vRANs) brings a new challenge with regards to computer architectures.
  It requires to select or design computing technologies that provide a sufficient level of performance
  while maximizing the flexibility and efficiency of the implemented networks.
  Several solutions addressing this challenge were proposed,
  relying on general purpose processors as well as hardware accelerators.
  This work describes our effort to enable an intensive vRAN deployment
  using the 5G software stack OpenAirInterface on top of these computer architectures.
  We had to adapt the software stack to leverage the capabilities of hardware
  and to find how to scale up the vRAN deployment with several vRAN instances sharing a server.
  We describe in this work our improvements to the stack and their effect on performance.
  We also share our observations on the behavior of the computer architectures
  and how they affect our deployment.
  We finally discuss the limitations of our deployment and further efforts
  to implement better vRAN deployments.
\end{abstract}

\section{Introduction}

\label{sec:introduction}

As the concept of virtual and open Radio Access Network (RAN) is flourishing,
silicon industrials such as Intel or AMD have been releasing several solutions
providing the computing capabilities necessary to efficiently implement
5G RANs at the same time as
addressing the requirements in term of flexibility and scalability of virtual RANs (vRANs).
Such solutions are already leveraged by major commercial network vendors
to implement their RAN or vRAN solutions~\cite{lightreading:chip-choices-kickstart-open-ran-war-between-lookaside-and-inline, fujitsu:whitepaper-inline-vs-lookaside}.
We will refer hereafter to this kind of architectures as {\bf vRAN computer architectures}.

Open-source 5G software stacks also turn to implementing vRANs
following the guidelines of the Open RAN (O-RAN) Alliance~\cite{o-ran:building-the-next-gen-ran, o-ran:architecture-description}.
FlexRAN~\cite{intel:flexran-reference-architecture, intel:flexran}
stands as a reference design for implementing the O-RAN specifications.
But other stacks also achieved end-to-end O-RAN compliant deployments:
OpenAirInterface~\cite{KALTENBERGER2025111410, oai:openairinterface5g}
or the srsRAN Project~\cite{10.1145/2980159.2980163, srs:srsRANProject}
--- which has recently been rebranded as OCUDU~\cite{linux-foundation:ocudu} ---.

Even though most of these 5G stacks are mainly aimed at research purposes,
some ambition to achieve commercial grade scenarios and performance.
This would enable research with commercial grade setups.
These stacks could also offer components for commercial implementations
thanks to the disaggregation of network architectures in the O-RAN standard.
Then, in addition of enabling the O-RAN network architectures and functionalities,
these 5G software stacks should as well leverage vRAN computer architectures
in order to approach the efficiency of commercial networks
in term of infrastructure and energy costs.

The effort we report within this work aims at enabling an open-source 5G software stack
to leverage three vRAN computer architectures from the industry
and to implement a deployment of vRANs sharing a single server with this stack.
One of our motivation through this effort is
to enable large-scale mobile network deployments
on top of a large-scale research infrastructure
for experimental purposes or for implementing a campus network.
For instance, the vRAN computer architectures we use are encountered in the
Scientific Large-scale Infrastructure for Computing/Communication Experimental Studies Research Infrastructure (SLICES-RI)~\cite{FDIDA2022189, slices-ri:slices},
a European research platform to support large-scale, experimental research focused on mobile networking,
supported by the European Strategy Forum on Research Infrastructures (ESFRI)~\cite{esfri:slices}.
They are also encountered on the Platforms for Advanced Wireless Research (PAWR)~\cite{nsf:pawr, BONATI2020107516} by the U.S. National Science Foundation.
The latter already offered testbeds for works involving vRAN computer architectures, for example Savannah~\cite{10.1145/3636534.3698843, 10.1145/3636534.3690707}.
A second motivation of this work is
to go one step forward towards enabling the use of 
open-source 5G software stacks as components within commercial solutions.

We chose to work with an open-source software stack widely adopted by the researcher's community
and to merge our contributions into the mainline code of the stack.
Then the vRAN deployment we achieve can be leveraged by a wide community
of researchers and developers.
OpenAirInterface stands as one of the most widely adopted stacks
and we have already been strongly involved in the development of the stack.
Then, the wide adoption and our knowledge of the stack and its environment
strongly incentivized us in pursuing our work with this stack.
Deployments of vRAN experimental setups using OpenAirInterface
are encountered in the literature~\cite{10620849}. 
But the OpenAirInterface stack has never been demonstrated in the literature 
on top of vRAN computer architectures.
It has never been demonstrated neither in an intensive deployment
involving many vRAN instances sharing a single server.

We demonstrate in this work that we enabled the OpenAirInterface stack to implement an intensive vRAN deployment
on top of some commercial computer architectures designed to implement the vRAN physical layer.
We observe the behavior of our deployment and identify necessary
improvements of the OpenAirInterface stack and of our setup to enable further and better scenarios.
As a first step, Section~\ref{sec:background}
gives elements of background on the vRAN architecture and its requirements for computing.
It also introduces three target systems, each featuring a different vRAN computer architecture.
Section~\ref{sec:enhancements} exposes the changes we contribute in the 
OpenAirInterface stack in order to enable it to leverage the three vRAN computer architectures
and to enhance the performance of the physical layer.
Section~\ref{sec:scale-up} introduces a challenging vRAN deployment
and explains how we manage to implement it on the three target systems.
The deployment consists in several 5G FR1 small-cell instances
implemented with OpenAirInterface and sharing a single host system.
This section also presents our observations on the behavior of the deployment.
Finally, section~\ref{sec:discussions}
concludes on the capabilities and limitations
of the OpenAirInterface stack with our contributions
and of our deployment.
We also discuss there further improvements that can be added to the stack
and use cases that are enabled by the current stack.

\section{Background}

\label{sec:background}

\subsection{vRAN}

\paragraph*{A new paradigm}
With regards to previous generations of mobile networks,
5G introduced new degrees of freedom in the use cases and architectures of Radio Access Networks (RAN).
5G introduced new kinds of connectivity use cases in addition to classical mobile broadband,
for example "massive Machine Type Communication" (mMTC) and "ultra-Reliable Low-Latency Communication" (uRLLC).
A high diversity of networks is also expected in term of size, range, performance or other capabilities,
especially with the advent of private mobile networks.
Operators were incentivized into resorting to disaggregated network architectures together
with bringing network functions to the cloud~\cite{china-mobile-research-institute:c-ran, ntt-docomo:5g-open-ran}
following the guidelines of the O-RAN Alliance~\cite{o-ran:building-the-next-gen-ran, o-ran:architecture-description}.
This disaggregation and cloudification not only enables to adapt network architectures to the new use cases
but also brings expenses savings for the operators by removing network vendors solutions lock-in
and leveraging the optimizations enabled by the cloud.
With these new degrees of freedom aroused a need in RAN solutions for flexibility, scalability
and ability to be brought to the cloud~\cite{BONATI2020107516}.
These solutions require from computer architectures to provide more flexibility and an ability to be shared.

Claims arose that past generation purpose-built hardware architectures
are not anymore fitting these new requirements and network architectures.
Designing new computer architectures leveraging the powerful general-purpose computing devices already in use for cloud computing or AI use cases
seems a straightforward path to implement flexible and scalable mobile network solutions~\cite{intel:whitepaper-vran-boost, amd:modernize-your-network}.
On the other hand, novel RAN computer architectures shall provide the performance level necessary
to seamlessly handle real-time and demanding baseband processing workloads.
Then, silicon industrials released a number of alternative new computer architectures
to implement the 5G RAN stack while virtualizing and sharing the infrastructure.
They rely on General Purpose Processors (GPP) as well as Hardware Accelerators (HA).
To be more specific, these architectures are designed for implementing the High-PHY layer
of the RAN in the O-RAN standard architecture.

\paragraph{vRAN High-PHY}
The O-RAN standard specifies the function and interfaces
of a number of network functions which make together the RAN.
The RAN protocol stack is scattered across these network functions.
The Radio Unit (RU) on the cell site implements the lower physical (Low-PHY) layer
while the Distributed Unit (DU) located on edge cloud locations close to the cell sites
implements the upper physical layer (High-PHY) and some higher layers.
The RAN protocol stack is therefore split at some determined point known as the 7.2 split
where a standard network interface known as the O-RAN 7.2 fronthaul~\cite{o-ran:fhi-72-cus, o-ran:fhi-72-m}
ensures the connection of the O-RAN DU and RU and of the two halves of the physical layer.
High-PHY corresponds to the part of the RAN protocol stack located
between the top of the physical layer and the 7.2 split.
It operates under tight real-time constraints
in order to respect the assignments of radio resources through time
that is determined by the higher Medium Access Control (MAC) layer
and to enable exchanges of transmissions, acknowledgement and retransmissions
required by the Hybrid Automatic Repeat reQuest (HARQ) protocol.
By seating in an edge location, the DU is the first network function up from the cell
that can be virtualized in an edge cloud infrastructure.
The DU can share the edge cloud with
other DUs, other user plane network functions and local end services.
It is for this reason that the efforts of implementers concentrated on
the DU and on the High-PHY layer which is the most compute intensive
part of the DU.

The High-PHY layer consists of a digital signal processing pipeline
with a downlink (DL) and an uplink (UL) direction.
Channel coding is a part of High-PHY.
It relies on Low-Density Parity-Check (LDPC) codes~\cite{doi:10.36227/techrxiv.174002467.71268212/v2}
to fix transmission errors introduced by the channel.
Its UL component, channel decoding is the dominant computing consumers in UL High-PHY.
Its DL counterpart, channel encoding is also one of the dominant computing consumer in DL High-PHY
together with precoding, which reckons DL data per antenna port to generate radio beams~\cite{10.1145/3386367.3431296, 10.1145/3452296.3472894}.
Of course, the balance between UL and DL traffic greatly influences the share
of computing consumed by the dominant UL and DL components in the whole High-PHY.
Other processing steps that can have a significant footprint are
channel estimation and compensation which attempt to compensate the effects of the channel by modeling it
and direct and inverse Fourier transforms which transform time domain signal into frequency domain data and vice-versa.

\subsection{The Architectures}

\paragraph*{The state of the art}
It is demonstrated that GPPs are able to handle even massive Multiple Input Multiple Output (MIMO)
broadband processing in the Frequency Range 1 (FR1) spectrum
and 2T2R MIMO in the Frequency Range 2 (FR2) spectrum with tighter time constraints~\cite{10.1145/3386367.3431296, 285088, 10.1145/3615453.3616521}.
On the other hand, HAs can also be integrated in a RAN while allowing hardware vendor exchangeability
and virtualization provided that HAs seamlessly interface with software through properly defined standard interfaces.
That is what the O-RAN standard provides by specifying its Accelerator Abstraction Layer (AAL)~\cite{o-ran:aal-ganp}.
This interface enables to offload operations with a defined profile to a Logical Processing Unit (LPU)
hiding one or many HAs or software libraries implementing the profile.
One commonly encountered implementation of the AAL is the BaseBand Device (BBDev) application
of the Data Plane Development Kit (DPDK) which acts as a performant real-time
driver for a variety of HAs and software libraries~\cite{dpdk:dpdk-doc-bbdev}.
We provide an overview of recent works on vRAN computer architectures with references
in section \ref{sec:related-works}.

\paragraph*{Our infrastructure}
We have three alternative computing architectures at our disposal
for implementing our vRAN deployment.
All of them were implemented by silicon industrials
and released as commercial products or as engineering samples.

\subsubsection*{1\textsuperscript{st} High-Performance Processor}
The simplest architecture relies on a general-purpose processor
with a high count of high-performance cores.
It is designated hereafter as High-Performance Processor
or with the acronym {\bf HPP}.
A higher core count as possible is privileged to leverage the economy of scale
by sharing the resource cost and energy consumption
between as many network instances as possible.
In this study we use a processor from the AMD EPYC 9005 series, an AMD EPYC 9575F,
with 64 Zen5 cores able to operate at clock frequencies up to 5 GHz and featuring AVX512.
This processor has 48 kilobytes of L1 data cache, 32 kilobytes of L1 instruction cache and 1 megabyte of L2 cache per single core
as well as 32 megabytes of L3 cache per group of 8 cores~\cite{amd:5th-gen-epyc-white-paper, amd:epyc-9005-series-architecture-overview}.
The system has 128 gigabytes of DDR5 memory configured with a transfer speed of 6000 megatransfers per second.
An Intel E810-C-Q2 Quad Small Form-factor Pluggable (QSFP) Network Interface Card (NIC)
is interfaced via a Peripheral Component Interconnect express (PCIe) gen4 x16 interface.
The system runs Ubuntu 25.04 with kernel 6.14.0-1011-realtime as host Operating System (OS).

\subsubsection*{2\textsuperscript{nd} Efficient General-Purpose Processor \& RFSoC}
The next candidate architecture features a HA for offloading channel coding with AAL.
The HA takes the form of an RF System-on-Chip (RFSoC) on a PCIe board
mounted in a processor-based host.
We have at our disposal a T2 Telco Accelerator Card from AMD~\cite{amd:t2-telco-accelerator-card}.
DPDK BBDev is used to interface the HA with software.
In  choosing the processor of the host, we try to leverage the fact that
the processor is relieved from channel coding by the RFSoC.
AMD offers two types of architecture within its 4\textsuperscript{th} generation of
EPYC processors, one series focuses on high-performance --- the 9004 series ---
with similar processors as the one we used for the HPP system,
and another series focuses on efficiency with regard to silicon and energy consumption --- the 8004 series ---.
We chose a processor from the latter series, an AMD EPYC 8534P,
with 64 Zen4c AMD64 cores reaching clock frequencies up to 3.1GHz and featuring AVX512.
This processor has 32 kilobytes of L1 data cache, 32 kilobytes of L1 instruction cache and 1 megabyte of L2 cache per single core
as well as 16 megabytes of L3 cache per group of 8 cores~\cite{amd:4th-gen-epyc-white-paper, amd:epyc-8004-series-architecture-overview}.
The system has 128 gigabytes of DDR5 memory configured with a transfer speed of 4800 megatransfers per second.
The T2 Telco Accelerator Card is interfaced via PCIe gen3 x16.
An Intel E810-C-Q2 QSFP NIC is interfaced via PCIe gen4 x16.
The system runs Ubuntu 25.04 with kernel 6.14.0-1011-realtime as host OS.
This solution is designated hereafter as Efficient Processor \& RFSoC
or with the acronyms {\bf EP-RFSoC}~\cite{lenovo-press:vran-du}.

\subsubsection*{3\textsuperscript{rd} vRAN Processor}
Another architecture competing with the one above offers to embed the HA in the package
of the host processor instead of having it on a separate RFSoC.
It saves the cost of an additional RFSoC even though the silicon manufacturer
may sell such a processor with this feature at a higher price than one without.
From a functional perspective, it removes the need of costly memory transfers
between the host and the RFSoC.
The HA does not have its own dedicated memory like an RFSoC on a PCIe board
but uses the main system memory mostly for data aggregation over repeated transmissions for channel decoding.
We have at our disposal an Intel processor from the Sapphire Rapids Edge Enhanced series,
an Intel Xeon Gold 6433N~\cite{intel:sapphire-rapids-edge-enhanced},
with 32 AMD64 cores reaching clock frequencies up to 2.7GHz and featuring AVX512
and an HA for channel decoding and encoding with the LDPC code.
The HA featured in this processor is marketed as "vRAN Boost".
DPDK BBDev is used to interface the HA with software~\cite{dpdk:dpdk-doc-vrb1}.
This processor has 48 kilobytes of L1 data cache, 32 kilobytes of L1 instruction cache and 2 megabytes of L2 cache per single core
as well as a 32 megabytes L3 cache shared by the 32 cores.
The system has 128 gigabytes of DDR5 memory configured with a transfer speed of 4400 megatransfers per second.
An Intel E810-C-Q2 QSFP NIC is interfaced via PCIe gen4 x16.
The system runs Red Hat Enterprise Linux 10.1 with kernel 6.12.0-124.35.1.el10\_1.x86\_64 as host OS.
The cores of the processor also feature the AVX512-FP16 instruction set extension
which provides instructions operating on 16 bits floating point numbers over 512 bits vector
in addition to the instructions already provided by the AVX512 extension.
This processor is designed to provide an all-in-one solution
for implementing the vRAN High-PHY.
For this reason, it is hereafter designated as vRAN Processor
or with the shorthand {\bf vRANP}.

\section{Enhancements of the OpenAirInterface RAN stack}

\label{sec:enhancements}

Leveraging the capabilities of the three architectures becomes possible
with enhancements to OpenAirInterface High-PHY
that we contributed during the past two years.
Some of our contributions were already merged in the mainstream stack
as of tag "2026.w06" from February 2026.
Other contributions are available as development branches
from the public code repository of the stack.
We aim to merge our remaining contributions
in the mainstream stack in the near future.

\subsection{LDPC channel coding}
As explained in section~\ref{sec:background},
channel coding with the LDPC code is the most
important computation consumer in High-PHY
UL and DL.
Then, a significant part of our work on the stack
tried to leverage the capabilities of the
vRAN computer architectures to implement
efficiently channel coding.
We contributed in enabling the offloading
of channel coding to an HA with AAL.
We also improved the efficiency of the  offloading
by reworking the integration of
channel coding libraries in the OpenAirInterface stack
and we finally optimized the
channel encoding implementation on GPPs.

\paragraph*{Offload to a HA with AAL}
We contributed in enabling
an AAL compliant offload of channel encoding and decoding with LDPC
relying on DPDK BBDev in the OpenAirInterface stack.
It is available as one shared object that can be optionally compiled
and loaded~\cite{oai:ldpc-offload-setup}.
As of the mainstream tag "2026.w06", 3 HAs are supported by the stack:
\begin{itemize}[noitemsep]
\item {\bf T2 telco accelerator}
      An RFSoC on a PCIe board from AMD
      featuring 8 Soft-Decision Forward Error Correction (SD-FEC) cores~\cite{amd:t2-telco-accelerator-card}.
      We have this accelerator in the EP-RFSoC system.
\item {\bf ACC 100}
      An accelerator from Intel.
      We do not own this accelerator
      but it is operated and tested by
      another contributor to the stack.
\item {\bf vRAN Boost}
      An accelerator from Intel, also known as ACC 200
      is embedded in the Intel Sapphire Rapids Edge Enhanced processors~\cite{intel:whitepaper-vran-boost, intel:sapphire-rapids-edge-enhanced, dpdk:dpdk-doc-vrb1}.
      We have it in the vRANP system.
\end{itemize}

BBDev leaves the possibility
to the HAs to implement some capabilities or not
in order to reflect
the actual capabilities of the hardware.
The integration of BBDev in the
OpenAirInterface stack had to
accommodate the respective capabilities
of the different HAs.

The main difference in capabilities we had to address
was that some HAs featured internal memories and others not.
The AMD T2 and Intel ACC 100 have an internal memory
for aggregating data across retransmissions
in order to fix transmission errors during channel decoding.
On the other hand, the Intel vRAN Boost which is embedded in a processor
does not have an internal memory.
Instead, aggregated data should be stored
in the memory of the host system.

Another, more surprising, difference between the HAs was
the support of Code Block and Transport Block interfaces.
In 5G LDPC coding, the Transport Blocks (TB) sent to or received from
each User Equipment (UE) within a slot are segmented into code segments also known as
Code Blocks (CBs) to be encoded or decoded~\cite{3gpp:38-212}.
BBDev offers interfaces to process both TBs and CBs, and
HAs are supposed to implement both interfaces.
In reality, not only the HAs do not implement the two interfaces,
but they do not implement the same interface in some cases.
While the AMD T2 implements only the CB interface in any case,
both Intel ACC 100 and vRAN Boost implement the CB interface in most cases
but users must use the TB interface when a TB breaks down in only one CB.

\paragraph*{implementation interface rework}
In order to enable an efficient offloading
of channel coding to HAs with BBDev,
another major contribution we made was to rework
the interface to LDPC coding implementations
in the OpenAirInterface stack.

Depending on the employed modulation and coding
and on the number of UEs scheduled on one slot,
the number of CBs --- or code segments --- processed
in one slot can vary from 1 to 132 for a 100MHz 4T4R RAN.
The interface of the channel coding implementations was previously
processing 1 CB per call to the decoder and 8 CBs per call to the encoder.
One call to either decoder or encoder was performing only LDPC coding
in the strict sense, which does not include rate matching.
This design was fitting the original GPP based implementation of OpenAirInterface.

But this interface was too limited to enable an efficient implementation with BBDev.
The interface for LDPC coding of BBDev includes LDPC coding in the strict sense
but also rate matching.
The reworked interface of LDPC coding includes code rate matching.
It also enables the processing of all the CBs of one slot in one call to either decoder or encoder.
Then, when the implementation is called,
BBDev has all the CBs of the slot at its disposal to
organize their offloading and processing in an efficient manner.

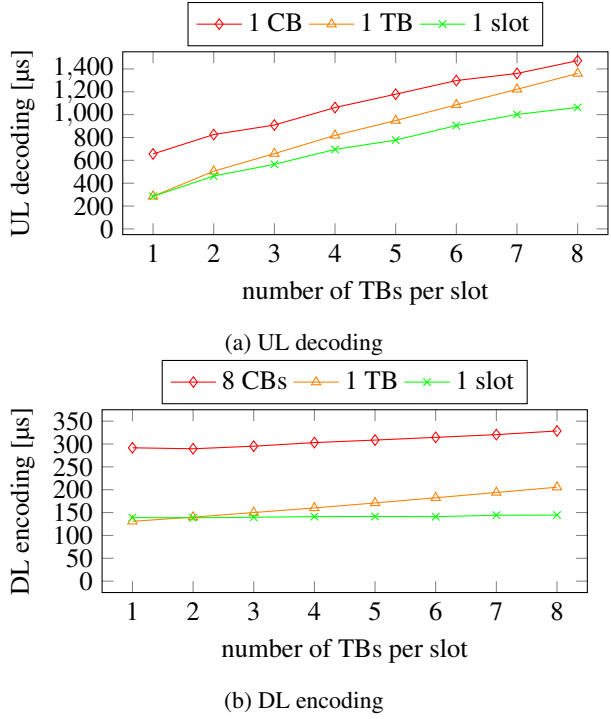
\begin{figure}
    \begin{subfigure}[][4.7cm][t]{0.45\textwidth}
        \begin{tikzpicture}
        \begin{axis}[
          xlabel={number of TBs per slot},
          ylabel={UL decoding [µs]},
          xmin=0.5, xmax=8.5,
          ymin=-50, ymax=1550,
          xtick={1,2,3,4,5,6,7,8},
          ytick={0,200,400,600,800,1000,1200,1400},
          legend style={
            at={(0.5,1.03)},
            anchor=south
          },
          legend columns=4,
          height=4cm,
        ]
        \addplot[
          color=red,
          mark=diamond,
        ]
        coordinates {(1,655.72)(2,825.73)(3,908.97)(4,1062.03)(5,1178.90)(6,1298.43)(7,1360.31)(8,1472.36)};
        \addlegendentry{1 CB}
        \addplot[
          color=orange,
          mark=triangle,
        ]
        coordinates {(1,285.23)(2,504.85)(3,658.90)(4,818.16)(5,948.10)(6,1085.88)(7,1221.70)(8,1360.34)};
        \addlegendentry{1 TB}
        \addplot[
          color=green,
          mark=x,
        ]
        coordinates {(1,284.73)(2,462.69)(3,564.82)(4,695.49)(5,777.32)(6,904.56)(7,1001.67)(8,1062.92)};
        \addlegendentry{1 slot}
        \end{axis}
        \end{tikzpicture}
        \caption{UL decoding}
        \label{fig:plot_multi_ue_dec}
    \end{subfigure}
    \begin{subfigure}[][4.7cm][t]{0.45\textwidth}
        \begin{tikzpicture}
        \begin{axis}[
          xlabel={number of TBs per slot},
          ylabel={DL encoding [µs]},
          xmin=0.5, xmax=8.5,
          ymin=-25, ymax=375,
          xtick={1,2,3,4,5,6,7,8},
          ytick={0,50,100,150,200,250,300,350},
          legend style={
            at={(0.5,1.03)},
            anchor=south
          },
          legend columns=4,
          height=4cm,
        ]
        \addplot[
          color=red,
          mark=diamond,
        ]
        coordinates {(1,291.70)(2,289.56)(3,295.20)(4,303.08)(5,308.60)(6,314.58)(7,320.46)(8,328.56)};
        \addlegendentry{8 CBs}
        \addplot[
          color=orange,
          mark=triangle,
        ]
        coordinates {(1,130.86)(2,139.96)(3,150.03)(4,160.12)(5,171.00)(6,182.34)(7,193.97)(8,205.44)};
        \addlegendentry{1 TB}
        \addplot[
          color=green,
          mark=x,
        ]
        coordinates {(1,138.90)(2,138.60)(3,139.77)(4,141.07)(5,141.44)(6,140.95)(7,144.21)(8,144.45)};
        \addlegendentry{1 slot}
        \end{axis}
        \end{tikzpicture}
        \caption{DL encoding}
        \label{fig:plot_multi_ue_enc}
    \end{subfigure}
    \caption{channel coding time on the EP-RFSoC system with 1 to 8 TBs per slot
             for different coding function interfaces}
\end{figure}

The gain in term of processing time obtained with this change of the LDPC coding implementations interface
is illustrated by \autoref{fig:plot_multi_ue_dec} and \autoref{fig:plot_multi_ue_enc}.
The figures show the average processing times for the channel decoding and encoding
of one slot in a 100MHz numerology 1 RAN operating
with Modulation and Coding Scheme (MCS) index 28 from MCS index table 1
that were obtained on the EP-RFSoC system.
There is one line for each interface to the LDPC coding implementation
that are encountered in the OpenAirInterface codebase history:
\begin{enumerate}[noitemsep]
  \item The former interface that processes 8 CBs per call for encoding and 1 CB per call for decoding
  \item An interface enabling to process one TB per function call
  \item The newer interface enabling to process the entire slot in one function call
\end{enumerate}
One slot breaks down into 26 CBs for 1 TB that occupies the whole slot.
The Physical Resource Blocks (PRBs) of the slot
are equally shared between 1 to 8 TBs.
Then these plots enable to observe the overhead of
breaking a slot down into multiple TBs with the three interfaces,
knowing that when a slot is shared between multiple UEs, each UE gets its own TB.
For both decoding and encoding, there is an improvement with the third interface compared to the first one.
The improvement remains whatever the number of TBs in the slot.
Meanwhile, with the second interface,
the improvement that is seen for one TB per slot decreases with the number of TBs in the slot.
Indeed, increasing the number of TBs increases the number of function calls required
even though the amount of data to process remains the same.
Then, we can conclude that the reworked interface enables
significant processing time reductions for channel encoding and decoding
compared to the former interface whatever the number of TBs per slot.
But we also notice that even with the new interface
the time to perform LDPC decoding significantly increases with the number of TBs.
This is a behavior we did not manage to explain as of writing these lines as
it was discovered while performing the present evaluation.

\paragraph*{encoder on GPPs}
In this process, we also rewrote the LDPC encoding and DL code rate matching on GPPs
to better leverage SIMD extensions.
SIMD instruction set extensions are sets of instructions on a processor
providing operations on bit vectors larger
than the usual word length of the processor.
Example of SIMD instruction set extensions 
optionally used in the OpenAirInterface stack are
AVX512 for up to 512 bits vectors on AMD64 systems
and Neon for up to 128 bits vectors on Arm64 systems.
This change enables for example a reduction of the processing time of up to 21\%
for channel encoding and DL code rate matching
in a 100MHZ 4T4R numerology 1 RAN on the HPP system
with 5 worker threads.
This improvement is merged in the mainstream stack as of tag "2026.w06".

\subsection{Precoding \& Resource Mapping}
Another improvement we contribute is to improve DL precoding and layer mapping in two ways.

First of all, we enabled precoding to leverage SIMD extensions.
This change enables for example a reduction of the processing time of up to 29\% for performing precoding and resource mapping
in a 100MHZ 4T4R numerology 1 RAN on the EP-RFSoC system with a single worker thread.

We added as well the option to execute resource mapping and precoding
on multiple worker threads while they are executed on only one main DL High-PHY thread in the mainstream stack.
This enhancement enables for example a processing time reduction
up to 67\% for performing precoding and resource mapping with 8 worker threads
in a 100MHZ 4T4R numerology 1 RAN on the EP-RFSoC system.

The leveraging of SIMD for precoding is merged in the mainstream OpenAirInterface stack as of tag "2026.w06".
The use of multiple worker threads for precoding and resource mapping
on the other hand is not merged yet in the mainstream stack.

\section{Scaling up the vRAN deployment}

\label{sec:scale-up}

We scaled up the number of vRAN instances
executed on each of the three alternative systems
to the maximum extent we could achieve
with our enhanced OpenAirInterface stack.
We proceeded by implementing the deployment
of many instances of
small-cell 5g base stations (gNB) under traffic load
on a pre-production setup.
We first determined the minimum set of resources
necessary to operate a gNB on each system.
Then, we replicated this gNB on each system up to the
maximum number of instances achievable while keeping
the gNB instances functional.

\subsection{Implemented deployment}

The deployment consists in the execution on one system
of several small-cell gNB instances
providing mobile broadband connectivity.
Such deployment can be used to implement
a campus network or a corporate private network
with computing hosted on premises or in an edge-cloud.
The small-cell gNB instances can also be regarded like the beams of a massive MIMO gNB
in which case this small-cell scenario turns to be analog in term of High-PHY computing to a massive MIMO scenario.
Our pre-production setup doesn't offer enough radio appliances to accommodate all the gNB instances.
Fortunately, the OpenAirInterface stack offers a "phy-test" mode
to stress the stack by emulating the connection of a User Equipment (UE).
This "phy-test" mode can be used together with
an O-RAN 7.2 fronthaul executing exactly as it would with an RU connected
but without a physical connection to an RU.
Then, each instance has an O-RAN 7.2 fronthaul,
but only one instance connects to a real RU~\cite{oai:oran-fhi72-tutorial}.
The instance with a real RU enables to assess the functionality with end-to-end connectivity tests.
The other instances execute in "phy-test" mode create a fair load on the system.
All the gNBs operate with 4 DL antennas and 4 UL antennas, a bandwidth of 100 MHz
on frequency band n77 and numerology 1, corresponding to a Transmission Time Interval (TTI) of 500 µs.
DL and UL traffic separation is enforced through Time Domain Division (TDD) and the TDD pattern is DDDSU.
This means that for each period of 5 slots, there are 3 slots dedicated to DL traffic,
one slot shared between DL and UL traffics and one slot dedicated to UL traffic.
On the gNB instances in "phy-test" mode, the emulated connection of a UE
has 4 layers and MCS index 27 from MCS index table 2 in the DL direction
and 2 layers and MCS index 16 from MCS index table 2 in the UL direction.
The emulated UE uses entirely all the DL and UL slots
but not the shared slot due to a limitation of the UE connection emulation.
Each gNB instance is executed within a dedicated Docker container.
Containerization enforces resources partitioning like processor cores partitioning.
The cores assigned to the gNB containers are isolated in the kernel parameters
so that they are not considered by the scheduler for scheduling other tasks than the gNB instances.
Some resources like network interfaces for O-RAN 7.2 fronthaul or HAs must be shared by the instances.
This is achieved through the use of Single Root Input \& Output Virtualization (SR-IOV)
and of the VFIO driver for Linux~\cite{dpdk:linux-drivers}
which enables to expose a single physical PCIe device through multiple virtual PCIe devices.
On the EP-RFSoC system only, it is necessary to execute an administration application
on a dedicated core isolated in the kernel parameters in order for multiple gNB instances to access the RFSoC.
Then each instance can use one of the virtual devices.

Starting from one deployed instance, we increase the number of instances one by one.
For each number of instances tested we check the functionality
of the gNB instance with a real RU connecting through end-to-end throughput tests.
We are checking
that the end-to-end DL throughput reaches 1200 Megabits per second
with 4 layers and a modulation order of 8
and that the end-to-end UL throughput reaches 90 Megabits per second
with 2 layers and a modulation order of 6.

\subsection{Resource Assignment}

\paragraph*{Resource requirements}
Sufficient resources have to be allocated to each gNB instance to execute successfully.
The OpenAirInterface documentation~\cite{oai:oran-fhi72-tutorial}
requests to assign some processor cores to given tasks.
These cores should be isolated, which means that the host operating system
will not schedule tasks on these cores other than the tasks we explicitly assigned.
The user should also avoid to assign these cores to other competing tasks.
We can summarize the respective roles of these 6 cores as:
\begin{itemize}[noitemsep]
\item 2 cores for DPDK worker threads for interfacing with the NIC and HA --- if any ---,
      namely the DPDK "IO" core and "worker" core.
\item 1 core for DPDK management, namely the DPDK "system" core.
\item 1 core for interfacing with the radio front-end,
      namely the "RU thread" core.
\item 2 cores dedicated to the High-PHY routines that do not require multi-threading,
      1 for DL and 1 for UL, namely the "L1 TX thread" and "L1 RX thread" cores.
\end{itemize}
The gNB also requires a set of cores not overlapping with the
6 statically assigned cores we just described.
This set of cores is known as the "thread pool".
It is a pool of cores to execute worker threads
implementing some of the most demanding
High-PHY workloads like channel coding
--- on the HPP system which does not offload channel coding to a HA ---,
precoding and resource mapping.
The size of this pool was of 7 cores for a 100MHz 4x4 FR1 gNB
in the materials we encountered on the OpenAirInterface repository~\cite{oai:fhi72-vvdn-docker-compose}.
Then, the requirements for cores of one gNB instance of our scenario
would be of 13 cores according to the materials
we had at our disposal.

For the interest of optimizing the processor resource usage by our deployment,
we reduced the number of required processors to 8 on the three target systems.
We managed this optimization by two means.
First of all, we observed that among the 6 cores statically assigned to single tasks.
The DPDK "system" core and "RU thread" had a very low level of usage.
We made two cores of the "thread pool" overlap with these
two assigned cores without introducing any issue.
We then also tried to reduce the size of the thread pool down to
the minimum size enabling the gNB to remain functional.
We managed to reduce the thread pool to 4 cores on the
three target systems.
Our final processor cores assignment was then:
\begin{itemize}[noitemsep]
\item 4 single task cores: DPDK "IO", DPDK "worker", "L1 TX thread" and "L1 RX thread" cores.
\item 2 cores: DPDK "system" and "RU thread" cores overlapping with the thread pool.
\item 2 cores for thread pool only.
\end{itemize}

\paragraph*{Computer architecture constraint}
Aside from the minimum resource required by one gNB instance,
another constraint we experienced in
assigning resources to gNB instances was coming
from the internal architecture of some processors.

The AMD EPYC 9005 and 8004 processors are manufactured from 8-cores core complexes with one L3 cache per core complex.
1 (EPYC 9005) or 2 (EPYC 8004) core-complexes are manufactured on a die and dies are assembled in a package to make the processor.
Communication between the dies and with other components is ensured by an additional IO die
~\cite{amd:5th-gen-epyc-white-paper, amd:epyc-9005-series-architecture-overview, amd:4th-gen-epyc-white-paper, amd:epyc-8004-series-architecture-overview}.
We observe a significant overhead on the gNB performance
when the range of the processor cores allocated to the gNB crosses the border of a core complex.
The behavior is worse when we cross the border between two dies.
In order to avoid the overhead and obtain optimal performances,
a gNB instance has to be contained on one core complex.

This behavior affects the HPP and EP-RFSoC systems.
On the vRANP system, we do not observe such a behavior.

\begin{figure}
    \begin{center}
        \includegraphics[width=0.8\columnwidth]{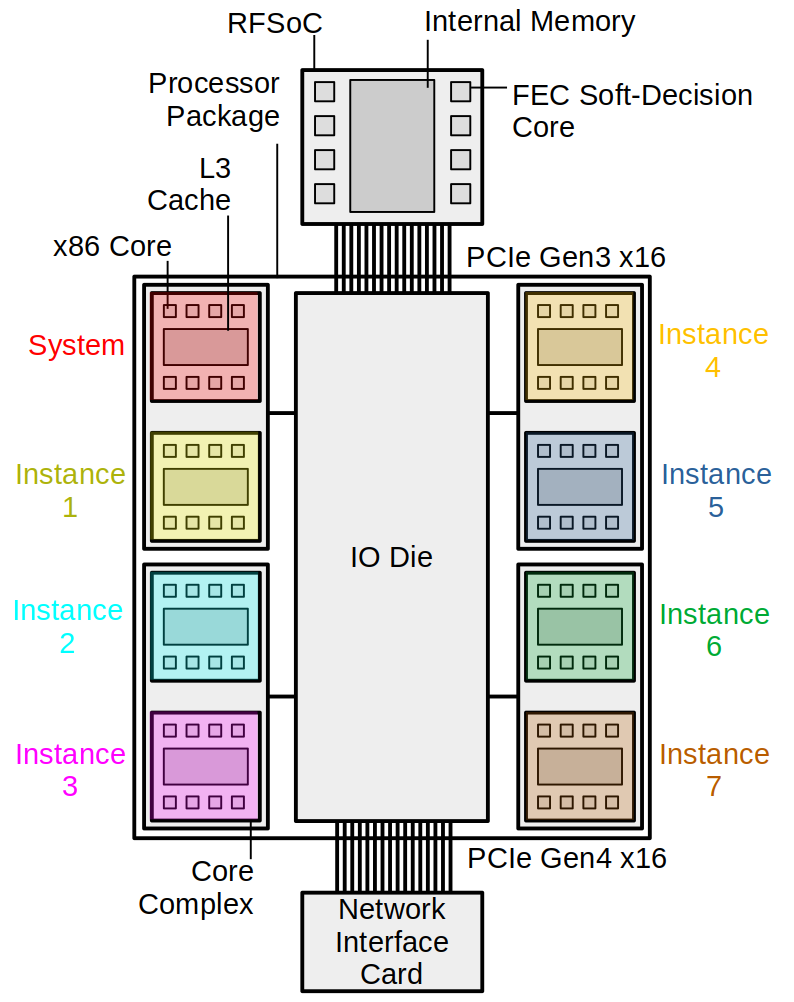}
    \end{center}
    \caption{Architecture of the EP-RFSoC system
             with envisioned gNB instances placement}
    \label{fig:arch_T2}
\end{figure}

\paragraph*{Assignment}
The processor of each system is partitioned so that each gNB instance is granted
the exclusive usage of a range of processor cores.

On the HPP system and EP-RFSoC system, one 8-cores core complex is assigned to each gNB instance.
except for one core complex that is left to the operating system.
We aim then to host 7 instances on these systems with their 64-cores processors.
\autoref{fig:arch_T2} gives an overview of the architecture of the EP-RFSoC system with the intended placement of the gNB instances.
The architecture and intended instance placement on the HPP system is very similar but with the difference that it doesn't feature any HA.

On the vRANP system, 8 cores are assigned to each gNB instance as well.
Then we aim to host up to 3 instances on this system with its 32-cores processor
with 8 cores left to the operating system.

Following the directives from the OpenAirInterface documentation,
we also set the processor core frequency to its maximum,
disable the idle states of the processor cores
and set their frequency governor to performance.

\subsection{Observations}

\begin{figure}[t]
    \begin{subfigure}[][6.8cm][t]{0.39\textwidth}
        \begin{tikzpicture}
        \begin{axis}[
          xmin=-20,
          xlabel={UL decoding [µs]},
          x tick label style={rotate=45},
          xtick={0,100,200,300,400,500,600,700,800},
          y tick label style={yshift=-0.8cm,rotate=40},
          ytick={1,2,3,4,5,6},
          yticklabels={
            {~~~~~~vRANP~3~inst.},
            {~~~~~~vRANP~1~inst.},
            {EP-RFSoC~7~inst.},
            {EP-RFSoC~1~inst.},
            {~~~~~~~~~~~~~~HPP~5~inst.},
            {~~~~~~~~~~~~~~HPP~1~inst.}
          },
        ]
        \addplot+ [
          boxplot prepared={
            lower whisker=282.997,
            lower quartile=284.930,
            median=288.689,
            upper quartile=296.289,
            upper whisker=326.284,
          },
          color=blue,
          style=solid,
          mark=otimes,
        ] table [row sep=\\,y index=0] { 259.734\\ 368.923\\ };
        \addplot+ [
          boxplot prepared={
            lower whisker=261.941,
            lower quartile=265.177,
            median=273.590,
            upper quartile=275.266,
            upper whisker=281.992,
          },
          color=blue,
          style=solid,
          mark=otimes,
        ] table [row sep=\\,y index=0] { 250.015\\ 328.316\\ };
        \addplot+ [
          boxplot prepared={
            lower whisker=159.329,
            lower quartile=165.439,
            median=176.856,
            upper quartile=199.219,
            upper whisker=231.378,
          },
          color=red,
          style=solid,
          mark=otimes,
        ] table [row sep=\\,y index=0] { 12.068\\ 496.515\\ };
        \addplot+ [
          boxplot prepared={
            lower whisker=149.765,
            lower quartile=152.128,
            median=154.962,
            upper quartile=158.207,
            upper whisker=175.754,
          },
          color=red,
          style=solid,
          mark=otimes,
        ] table [row sep=\\,y index=0] { 142.464\\ 200.241\\ };
        \addplot+ [
          boxplot prepared={
            lower whisker=393.100,
            lower quartile=450.130,
            median=505.730,
            upper quartile=566.991,
            upper whisker=636.321,
          },
          color=black,
          style=solid,
          mark=otimes,
        ] table [row sep=\\,y index=0] { 37.470\\ 791.591\\ };
        \addplot+ [
          boxplot prepared={
            lower whisker=403.331,
            lower quartile=430.670,
            median=485.961,
            upper quartile=537.670,
            upper whisker=615.221,
          },
          color=black,
          style=solid,
          mark=otimes,
        ] table [row sep=\\,y index=0] { 303.380\\ 735.871\\ };
        \end{axis}
        \end{tikzpicture}
        \caption{UL decoding}
        \label{fig:boxplot_dec}
    \end{subfigure}
    \begin{subfigure}[][6.8cm][b]{0.39\textwidth}
        \begin{tikzpicture}
        \begin{axis}[
          xlabel={DL encoding [µs]},
          x tick label style={rotate=45},
          xtick={0,50,100,150,200,250,300,350,400,450,500},
          y tick label style={yshift=-0.8cm,rotate=40},
          ytick={1,2,3,4,5,6},
          yticklabels={
            {~~~~~~vRANP~3~inst.},
            {~~~~~~vRANP~1~inst.},
            {EP-RFSoC~7~inst.},
            {EP-RFSoC~1~inst.},
            {~~~~~~~~~~~~~~HPP~5~inst.},
            {~~~~~~~~~~~~~~HPP~1~inst.}
          },
        ]
        \addplot+ [
          boxplot prepared={
            lower whisker=65.127,
            lower quartile=100.288,
            median=126.999,
            upper quartile=131.024,
            upper whisker=136.366,
          },
          color=blue,
          style=solid,
          mark=otimes,
        ] table [row sep=\\,y index=0] { 20.940\\ 153.838\\ };
        \addplot+ [
          boxplot prepared={
            lower whisker=102.019,
            lower quartile=108.566,
            median=110.658,
            upper quartile=112.460,
            upper whisker=114.272,
          },
          color=blue,
          style=solid,
          mark=otimes,
        ] table [row sep=\\,y index=0] { 7.191\\ 118.827\\ };
        \addplot+ [
          boxplot prepared={
            lower whisker=130.135,
            lower quartile=133.681,
            median=139.189,
            upper quartile=158.157,
            upper whisker=189.965,
          },
          color=red,
          style=solid,
          mark=otimes,
        ] table [row sep=\\,y index=0] { 11.848\\ 498.408\\ };
        \addplot+ [
          boxplot prepared={
            lower whisker=125.438,
            lower quartile=126.110,
            median=126.931,
            upper quartile=127.772,
            upper whisker=128.823,
          },
          color=red,
          style=solid,
          mark=otimes,
        ] table [row sep=\\,y index=0] { 38.017\\ 176.545\\ };
        \addplot+ [
          boxplot prepared={
            lower whisker=60.450,
            lower quartile=99.440,
            median=110.250,
            upper quartile=112.680,
            upper whisker=118.190,
          },
          color=black,
          style=solid,
          mark=otimes,
        ] table [row sep=\\,y index=0] { 22.010\\ 148.660\\ };
        \addplot+ [
          boxplot prepared={
            lower whisker=71.890,
            lower quartile=100.120,
            median=101.269,
            upper quartile=103.740,
            upper whisker=108.090,
          },
          color=black,
          style=solid,
          mark=otimes,
        ] table [row sep=\\,y index=0] { 24.890\\ 124.820\\ };
        \end{axis}
        \end{tikzpicture}
        \caption{DL encoding}
        \label{fig:boxplot_enc}
    \end{subfigure}
    \caption{Distribution of time for channel coding of a slot,
             whiskers are first and last decile, fliers are extrema}
\end{figure}

\begin{figure}[t]
    \begin{subfigure}[][6.8cm][t]{0.39\textwidth}
        \begin{tikzpicture}
        \begin{axis}[
          xmin=-50,
          xlabel={High-PHY UL [µs]},
          x tick label style={rotate=35},
          xtick={0,250,500,750,1000,1250,1500,1750,2000,2250},
          y tick label style={yshift=-0.8cm,rotate=40},
          ytick={1,2,3,4,5,6},
          yticklabels={
            {~~~~~~vRANP~3~inst.},
            {~~~~~~vRANP~1~inst.},
            {EP-RFSoC~7~inst.},
            {EP-RFSoC~1~inst.},
            {~~~~~~~~~~~~~~HPP~5~inst.},
            {~~~~~~~~~~~~~~HPP~1~inst.}
          },
        ]
        \addplot+ [
          boxplot prepared={
            lower whisker=1896.637,
            lower quartile=1921.196,
            median=1953.499,
            upper quartile=2085.933,
            upper whisker=2108.219,
          },
          color=blue,
          style=solid,
          mark=otimes,
        ] table [row sep=\\,y index=0] { 362.774\\ 2309.559\\ };
        \addplot+ [
          boxplot prepared={
            lower whisker=1874.875,
            lower quartile=1900.296,
            median=1918.903,
            upper quartile=2002.173,
            upper whisker=2022.937,
          },
          color=blue,
          style=solid,
          mark=otimes,
        ] table [row sep=\\,y index=0] { 342.521\\ 2257.361\\ };
        \addplot+ [
          boxplot prepared={
            lower whisker=1779.742,
            lower quartile=1802.906,
            median=1850.748,
            upper quartile=1950.908,
            upper whisker=2002.536,
          },
          color=red,
          style=solid,
          mark=otimes,
        ] table [row sep=\\,y index=0] { 340.811\\ 2225.380\\ };
        \addplot+ [
          boxplot prepared={
            lower whisker=1638.559,
            lower quartile=1645.610,
            median=1660.872,
            upper quartile=1832.060,
            upper whisker=1851.259,
          },
          color=red,
          style=solid,
          mark=otimes,
        ] table [row sep=\\,y index=0] { 331.016\\ 1941.805\\ };
        \addplot+ [
          boxplot prepared={
            lower whisker=1222.191,
            lower quartile=1353.312,
            median=1422.202,
            upper quartile=1516.562,
            upper whisker=1591.692,
          },
          color=black,
          style=solid,
          mark=otimes,
        ] table [row sep=\\,y index=0] { 198.010\\ 1787.903\\ };
        \addplot+ [
          boxplot prepared={
            lower whisker=1154.911,
            lower quartile=1257.182,
            median=1321.022,
            upper quartile=1406.451,
            upper whisker=1468.131,
          },
          color=black,
          style=solid,
          mark=otimes,
        ] table [row sep=\\,y index=0] { 183.381\\ 1630.542\\ };
        \end{axis}
        \end{tikzpicture}
        \caption{UL}
        \label{fig:boxplot_rx}
    \end{subfigure}
    \begin{subfigure}[][6.8cm][b]{0.39\textwidth}
        \begin{tikzpicture}
        \begin{axis}[
          xmin=-50,
          xlabel={High-PHY DL [µs]},
          x tick label style={rotate=35},
          xtick={0,200,400,600,800,1000,1200,1400,1600,1800,2000},
          y tick label style={yshift=-0.8cm,rotate=40},
          ytick={1,2,3,4,5,6},
          yticklabels={
            {~~~~~~vRANP~3~inst.},
            {~~~~~~vRANP~1~inst.},
            {EP-RFSoC~7~inst.},
            {EP-RFSoC~1~inst.},
            {~~~~~~~~~~~~~~HPP~5~inst.},
            {~~~~~~~~~~~~~~HPP~1~inst.}
          },
        ]
        \addplot+ [
          boxplot prepared={
            lower whisker=491.078,
            lower quartile=573.694,
            median=617.495,
            upper quartile=636.912,
            upper whisker=661.103,
          },
          color=blue,
          style=solid,
          mark=otimes,
        ] table [row sep=\\,y index=0] { 387.931\\ 1229.210\\ };
        \addplot+ [
          boxplot prepared={
            lower whisker=532.033,
            lower quartile=553.182,
            median=564.859,
            upper quartile=587.859,
            upper whisker=609.560,
          },
          color=blue,
          style=solid,
          mark=otimes,
        ] table [row sep=\\,y index=0] { 227.510\\ 1303.547\\ };
        \addplot+ [
          boxplot prepared={
            lower whisker=573.611,
            lower quartile=595.143,
            median=612.078,
            upper quartile=632.980,
            upper whisker=671.067,
          },
          color=red,
          style=solid,
          mark=otimes,
        ] table [row sep=\\,y index=0] { 246.890\\ 1920.532\\ };
        \addplot+ [
          boxplot prepared={
            lower whisker=527.11,
            lower quartile=535.574,
            median=544.787,
            upper quartile=555.063,
            upper whisker=567.422,
          },
          color=red,
          style=solid,
          mark=otimes,
        ] table [row sep=\\,y index=0] { 297.827\\ 1175.013\\ };
        \addplot+ [
          boxplot prepared={
            lower whisker=245.110,
            lower quartile=310.390,
            median=330.609,
            upper quartile=347.389,
            upper whisker=379.770,
          },
          color=black,
          style=solid,
          mark=otimes,
        ] table [row sep=\\,y index=0] { 155.290\\ 886.179\\ };
        \addplot+ [
          boxplot prepared={
            lower whisker=242.040,
            lower quartile=287.560,
            median=294.810,
            upper quartile=307.860,
            upper whisker=329.870,
          },
          color=black,
          style=solid,
          mark=otimes,
        ] table [row sep=\\,y index=0] { 126.970\\ 906.679\\ };
        \end{axis}
        \end{tikzpicture}
        \caption{DL}
        \label{fig:boxplot_tx}
    \end{subfigure}
    \caption{Distribution of High-PHY processing time of a slot,
             whiskers are first and last decile, fliers are extrema}
\end{figure}

On the three systems, as the number of instances sharing the system increases,
the behavior of individual instances evolves in several aspects.

\paragraph*{Functionality}
The change of behavior with the increasing number of instances
makes that we cannot deploy the number of gNB instances we expected on
the HPP system.
On this system, gNB instances manage to execute
for a maximum of 5 instances instead of the 7 expected.
When attempting to execute 6 gNB instances
some of the instances stop executing
before any throughput test is started.
The logs of the gNB instances that stop executed
report internal time deadline infringements.
OpenAirInterface ultimately asserts
on reaching a faulty state.
This behavior may be explained by the drop
of the processor clock frequency coming
with a higher usage.
Indeed, from around 4900MHz when no gNB instance is running,
the processor clock frequency drops around 4660MHz
when launching 1 gNB instance and 4260MHz when
5 instances are launched.

On the other hand,
The EP-RFSoC system
does execute successfully
and passes the throughput tests
with 7 instances out of the 7 expected.
The vRANP system as well
is able to pass the UL and DL throughput test
with the expected number of instances, 3 out of 3.

\paragraph*{Channel coding}
\autoref{fig:boxplot_dec} and \autoref{fig:boxplot_enc} show, for the three systems,
the distribution of times for performing the LDPC encoding and decoding of a slot
recorded during one second on the gNB with real radio
under the DL throughput test for encoding and UL throughput tests for decoding
for only one gNB instance and the maximum number of instances
that can be executed on each system.

For UL channel decoding the HAs clearly provide outstanding
performance in any situation
while a high-performance processor struggles against this task
for 1 instance as well as the maximum number of instances.

For DL channel encoding on the other hand,
the HAs are slightly outperformed by the high-performance processor,
but the encoding time remains close between the systems
for most points of the distribution
and for 1 instance as well as the maximum number of instances.

Beyond these general tendencies, our attention is drawn
on the EP-RFSoC system which shows
different pictures in the case with 1 instance
and in the case with 7 instances.
On every system, the case with the maximum number of instances
displays processing times that increased compared to the case with only one instance
for both UL and DL and for most point of the distribution.
This increase is rather limited compared to the absolute values
except for the maximum of the processing times
on the EP-RFSoC system which significantly increases
in the case of 7 instances compared to the case of 1 instance.
The maximum UL decoding time jumps from 200µs to 497µs (+149\%)
while the median only increases from 155µs to 177µs (+14\%).
The maximum DL encoding time jumps from 177µs to 498µs (+181\%)
while the median only increases from 127µs to 139µs (+9\%).
The 9\textsuperscript{th} decile also increases
more than other points of the distribution but less than the maximum.
It proves the occurrence of spikes of the processing time far ahead of common processing times.
These spikes may threaten the respect of real-time deadline when they occur.

We actually expected this behavior and we monitored the distribution of
channel coding times to know whether it would occur.
This is the result of contention between gNB instances for using the RFSoC.
It is a well-known issue of shared accelerators and a well-studied topic
especially when it comes to solutions to mitigate the risk
of real-time deadline infringement that this issue creates.
Recent studies on this topic are referenced in section~\ref{sec:related-works}.

Our deployment not only shows this behavior in an end-to-end deployment
but it also shows that this behavior does not always appear, as we
do not observe the behavior on the vRANP system which relies as well
on a HA for channel coding.
This difference between the EP-RFSoC and vRANP systems may be explained
by differences in the design of their respective HAs and by the different number of gNB instances.
But we cannot verify this hypothesis as we are missing information
on the design and capacity of the Intel vRAN Boost accelerator.
Regarding the T2 accelerator, we notice that the behavior occurs besides
its 8 SD-FEC cores and its announced supported throughputs of
35 Gigabits per second in DL and 12 Gigabits per second in UL which is far above the 
uncoded data throughput obtained in DL and UL by 7 gNB instances.
The summed theoretical uncoded throughputs of the 7 gNB instances reaches
around 12 Gigabits per second for DL and less than 1 Gigabits per second for UL.

Another question that arises when we observe the behavior of the EP-RFSoC system
is whether the extra time overhead has a significant impact on the total
processing time of the DL High-PHY layer which comprises much more tasks
than channel encoding.

\paragraph*{Global High-PHY computing}
There are several observations that can be made about the
processing times spent in the whole High-PHY in the UL and DL directions.
\autoref{fig:boxplot_rx} and \autoref{fig:boxplot_tx} show the distribution of the total times
for processing a slot in the High-PHY layer of the gNB with real radio
with 1 instance and the maximum number of instances under the UL and DL throughput tests.

In the UL direction, we observe that the HPP system shows in every points of the distribution lower processing times
than the EP-RFSoC and vRANP systems while the order was reversed when looking at channel coding alone.
As expected, the high-performance Zen 5 processor of the HPP system completes faster than the
less performant processors of the EP-RFSoC and vRANP systems all the tasks of
the UL High-PHY stack that are not handled by a HA.

In the DL direction, we observe as well an advance of the HPP system.

We can also quantify the effect on total High-PHY processing times of
the RFSoC sharing with 7 gNB instances on the EP-RFSoC system.
The maximum of the UL High-PHY processing time increased
from 1942µs for 1 instance to 2225µs for 7 instances (+15\%)
while the median increased from 1661µs to 1851µs (+11\%).
Then the effect of the UL decoding time spikes is rather
limited compared to the global increase of UL High-PHY processing time.
The increases are rather limited in comparison to the values for 1 instance.
The maximum of the DL High-PHY processing time increased
from 1175µs for 1 instance to 1921µs for 7 instances (+63\%)
while the median increased from 545µs to 612µs (+12\%).
Then DL High-PHY processing times
experience significant spikes that can represent a real
threat for the real-time compliance
of the DL High-PHY routine when they happen.
For a mobile broadband use case, the impact of this behavior
is not critical since a high reliability of the network is not required.
That would be different if our purpose was to implement
ultra reliable low-latency networks.
Then, a deeper study on the impact of the behavior on the reliability
and solutions to the issue would be relevant.

\paragraph*{Computing resource usage}
The High-PHY processing times show
a superiority of the HPP system with regards to the other systems
when it comes to providing guarantees on routines completion times
and therefore on the reliability and latency of the network.
On the other hand, the gNB instances on the
EP-RFSoC and vRANP systems
behave satisfactorily with regards to the
mobile broadband performance target we set.
In addition, the level of usage of the system resources
is better on the EP-RFSoC and vRANP systems,
on which we managed to execute the number of instances
we initially expected, than on the HPP system
on which 16 cores remain unused with only 5 instances
actually deployed over 7 expected.
Unfortunately, the 56 processor cores that we expected
to use for implementing 7 gNB instances cannot be
redistributed to implement 6 gNBs instead as this
would require from the set of cores allocated to each instance
to cross the die borders.
The 16 cores remaining unused may eventually be reused for
other best effort workloads if and only if
they do not excessively affect the behavior of the gNB instances.

The HPP has then a significant disadvantage
in comparison with the EP-RFSoC and vRANP systems
when it comes to the infrastructure investment per vRAN instance,
even though the final picture must also take into account
the respective price for affording the three systems.
Beyond this observation on infrastructure costs which belong to the
Capital Expenditures (CAPEX) for implementing the networks of our deployment,
we can observe the energy consumption of the instances
which is part of the Operational Expenditures (OPEX).

\paragraph*{Energy efficiency}
The provision of energy is indeed the most significant part of the OPEX for RANs with sometime up to 40\% of the OPEX
according to figures from the China Mobile Research Institute~\cite{china-mobile-research-institute:c-ran}.
We are able to measure the intake electrical power at the power supply on the three systems
through a standard Redfish~\cite{dmtf:redfish} management interface.
We measure this consumption when the efficiency of the system and loads are at their peak,
with the maximum number of gNB instances deployed on each system and throughput tests ongoing.

The total consumption of the HPP system with 5 gNB instances is 534W which is almost 107W per instance.
Even with the 7 instances initially expected, the consumption per instance would not have been lower than 76W on this system.
Meanwhile, the power consumption is only 309W for the EP-RFSoC system with 7 gNB instances which is almost 45W per instance
and 268W for the vRANP which is almost 90W per instance with 3 gNB instances.
Then the use of targeted acceleration with an efficient GPP shows its superiority in term of energy efficiency
compared to a high performance GPP.

This validates the choice to associate the E-GPP to the RFSoC for one of the systems to obtain a well calibrated computing capability
for a reduced power consumption while using a HPP with or without HA would be over-provisioned with a sub-optimal power draw.
Still, using a HPP may be relevant for implementing RANs with requirements on their reliability like with the uRLLC quality of service
or with higher numerologies like in the FR2 spectrum which implies tighter time constraints.

\section{Discussions}

\label{sec:discussions}

Through enhancing the OpenAirInterface stack,
achieving the pre-production deployment and observing,
we raised some subjects for discussions.
First of all, we will draw the status of the development
of our deployment: what was achieved and what are the remaining efforts.
Then we will discuss the limitations of our deployment
and of the OpenAirInterface stack as well as ideas to bridge the gaps.
Finally, we will reflect on the usage that can be made
of our deployment in the scope of large-scale experiments and beyond.

\subsection{Status of the deployment}

\paragraph{Achievements}
With the pre-production deployment we achieved, we addressed
some of the main technical challenges in order to implement
an efficient deployment of vRANs using OpenAirInterface
which were: 
\begin{itemize}[noitemsep]
\item To modify the OpenAirInterface stack with necessary changes
      to use it in an intensive deployment.
      It means mostly to support and use efficiently
      the targeted computer architectures.
\item To configure the stack in a way that optimizes
      its usage of the computing resources.
\item To scale up the deployment of network instances.
\item To record performance indicators that
      enable to compare different alternative
      computer architectures for our deployment
      and identify potential issues.
\end{itemize}

\paragraph{Upcoming Efforts}
Besides the challenges that were already addressed,
there are still some efforts to bring our deployment to production.
The main efforts we foresee is to design the
orchestration of the vRAN instances.
The orchestration should be aware of the specificities
of our deployment like the use of SR-IOV for interfacing with NICs and HAs or
the processor cores assignment constraints
especially on the AMD EPYC systems.
For instance, if we wish to deploy our work
on top of the SLICES Research Infrastructure
we have to find how the specificities of our deployment
can be taken into account in the
SLICES-RI blueprints~\cite{10620849, slices-sc:blueprints}.

\subsection{Limitations \& Enhancements}

\paragraph*{Limitation with regards to true vRAN}
The resource assignment constraints from the OpenAirInterface stack
and of the AMD EPYC processor architectures
prevent to leave gNB instances running concurrently
in a wide shared pool of processor cores
in our deployment as it should be in a vRAN scenario.
This is a major divergence from a fundamental principle of virtualized RAN
as the requirements for dedicated processor resources prevent
from sharing these resources concurrently with other vRAN instances or
even with other best-effort tasks.

Cores dedicated to single tasks are a common resort in real-time applications
in order to avoid any contention that may lead to missing completion deadlines.
On the other hand, solutions have been proposed to address
the problem of guaranteeing the real-time compliance of a RAN stack
on top of shared computer resources.
In the same fashion as a Real-Time Operating Systems (RTOS),
a vRAN aware scheduler interleaves tasks in a way
that vRANs do not lack the computing resources they need
at the moment they need them.
We provide an overview of these solutions with references
in section~\ref{sec:related-works}.

\paragraph*{Further Enhancements}
Some features are missing in the OpenAirInterface stack that would unlock new perspectives.

{\bf AMD Versal}
The T2 telco accelerator that we use in the EP-RFSoC system was discontinued by AMD.
Future computer architectures would rather rely on the newer AMD Versal RF Seriers SoCs.
Integrating this new generation of RFSoCs not only requires to integrate a new BBDev driver
--- if it is released ---,
but it also requires to accommodate to the fact that the Versal RFSoCs feature only LDPC decoding accelerators
and no LDPC encoding accelerators~\cite{amd:versal-rf, amd:versal-rf-brief}.
Then, LDPC encoding should be implemented on the host system processor.
This renews the question of the most appropriate architecture for the processor
between the dense and efficient Zen5c architecture and the high-performance Zen5 architecture~\cite{amd:5th-gen-epyc-white-paper, amd:epyc-9005-series-architecture-overview}.

{\bf vRANP support}
In order to fully leverage the capabilities of the vRANP architecture,
the stack should be enabled to use the Fast Fourier Transform (FFT) profile of the "vRAN Boost" HA
in order to accelerate the inverse FFT that is performed in High-PHY for the Physical Random-Access CHannel (PRACH).
The software should also be enabled to use
the AVX512-FP16~\cite{intel:whitepaper-vran-boost, intel:intrinsics-guide-avx512-fp16} instruction set extension.
This extension provides SIMD instructions to operate on 16 bits floating point samples within 512 bits vectors.
This enables the efficient processing of 16 bits floating-point complex symbols
--- meaning 16 bits for the real part and 16 bits for the imaginary part ---
which is the format of data used in the lower High-PHY layer between modulation and the 7.2 split.

{\bf Arm processor \& GPU}
Architectures offered by Nvidia relying on an Arm64 processor accelerated with
a GPU stand as an alternative to AMD64 based architectures upon which most 5G software stacks rely.
One hope behind such an architecture is to collocate vRAN and best-effort Machine Learning (ML) workloads
in order to maximize the usage of edge GPU infrastructures~\cite{9605055, 10621380}.
Nvidia demonstrates this possibility through its own PHY layer implementation called Aerial~\cite{nvidia:aerial}.
Mainstream OpenAirInterface does not feature any maintained capability of offload to a GPU.
We are working on enabling the stack to leverage this kind of architectures.
As of writing theses line, our work is not mature enough to match the network performance achieved with the other architectures.


{\bf O-RAN AAL integration}
The integration of the O-RAN AAL into OpenAirInterface is used only to
offload channel coding to a single HA while the capabilities
of the AAL as described in the standard are wider than this simple case~\cite{o-ran:aal-ganp}.
The standard leaves the freedom to implement many different operations or "profiles"
with, for each profile, queues of operations which can distribute to many implementations, either HAs or software libraries.
The ability of the AAL to distribute computing between a HA and a software library
can be used to reinforce the reliability of vRAN on a shared computing infrastructure by providing
variable computing resource in real-time as it has already been demonstrated with CloudRIC~\cite{10.1145/3636534.3649381}.


\paragraph*{Networking scenarios}
Our deployment involved FR1 gNB instances with numerology 1 which are not as challenging
in term of real-time constraints as FR2 gNBs with numerology 3 with completion deadlines 4 times shorter.
The FR2 use case was implemented by Savannah~\cite{10.1145/3636534.3698843, 10.1145/3636534.3690707} using a HA through the AAL interface
but without considering to share the computing infrastructure.
The OpenAirInterface stack supports the FR2 spectrum with numerology 3 and the 7.2 split.
Then we could consider as a future work to design a deployment of a FR2 or mixed networking scenario.

Our work doesn't enable neither the deployment of massive MIMO RANs.
Massive MIMO configurations are currently not achievable end-to-end
with OpenAirInterface which at best implements 8T8R MIMO
with two 4T4R O-RAN compliant RUs.
Enabling Massive MIMO with the stack requires to implement digital beamforming for Massive MIMO.
Massive MIMO RUs supporting the O-RAN 7.2 fronthaul are also hard to find.
The enabling of massive MIMO would unlock a new space for enhancements in OpenAirInterface High-PHY
in order to address the associated computing requirements.

\subsection{Use Cases}

When it is deployed on the target large-scale research infrastructure,
our deployment enables a few use cases.
We could also imagine further use cases beyond a use on the research infrastructure.

{\bf Large-scale experiments}
The most important use case is to enable
experiments at scale that would not be
achievable otherwise.
Works on efficient implementations of vRANs
can be tested in a setup reflecting
production operation conditions
based on our work.
Mobility experiments or data collection
in operational conditions
can be performed within campus networks
with a dozen of cells.
This can be enabled with
only one or two edge-computing nodes
based on our work.

{\bf Experiments on computer architectures}
Our work also enables further experiments with computer architectures.
Future experiments could extend our work by
implementing solutions to guarantee a real-time behavior with shared HAs
or enable sharing of processor cores.
Future experiments could as well introduce new architectures or concepts
to implement efficient vRANs operation.
Our work would then serve as a framework
or as a reference for comparison.

{\bf Small-cells deployment}
Finally, although this deployment was primarily designed for
a research infrastructure, we shall not overlook
potential commercial exploitations.
Our work indeed enables deployments of 100MHz 4x4 FR1 small-cells
providing mobile broadband connectivity
with an O-RAN architecture and a power consumption of 45W per DU.

\section{Related works}

\label{sec:related-works}

Many works try to find a hardware architecture
satisfying the requirements for performance and flexibility of vRANs.
Below, we provide an overview of recent developments on the topic.

{\bf Capable architectures}
Some works try to demonstrate that GPPs are able to handle massive MIMO baseband processing
without the assistance of any HA.
Agora~\cite{10.1145/3386367.3431296} demonstrates the feasibility of handling 64T64R FR1 numerology 0 with a single GPP based server.
It is extended by Hydra~\cite{285088} which handles up to 128T32R MIMO
and Agora-UHD~\cite{10.1145/3615453.3616521} for handling 2T2R FR2 numerology 3 with tighter completion deadlines.
But Hydra resorts to massive pools of processor cores spanning over multiple servers
for a single instance of the most massive MIMO configurations.
Agora for 2T2R FR2 numerology 3 on its end is outperformed by resorting to an HA as demonstrated with Savannah~\cite{10.1145/3636534.3698843, 10.1145/3636534.3690707}.
Savannah relies on the FlexRAN~\cite{intel:flexran-reference-architecture, intel:flexran} stack
which implements the standard O-RAN AAL interface.

{\bf Processor sharing}
Nevertheless, GPPs enable some optimization to increase their efficiency when they are not used at their full potential by a vRAN.
One optimization consists in collocating best-effort workloads other than the vRAN on the same cores provided that
a control loop is implemented to modulate the share of resources provided to the best-effort workloads
while guaranteeing to the vRAN sufficient resources to respect its real-time constraints as offered with Concordia~\cite{10.1145/3452296.3472894}.
Another optimization consists in modulating the processor frequency depending on cellular workloads
with an appropriate control system that guarantees the compliance with real-time constraints as offered with RENC~\cite{10.5555/3767955.3768020}.

{\bf HA sharing}
Sharing a HA while guaranteeing a real-time behavior is a challenge as we experienced in section~\ref{sec:scale-up}.
But HAs sharing cannot be overlooked without degrading the energy and resource usage efficiency in comparison to GPPs.
One way to walk around this issue is to share an HA between a vRAN and other best-effort workloads.
This was considered especially for GPUs~\cite{9605055}.
In the same fashion as for GPPs, the share of resources provided to the best effort workloads is modulated
in order to always guarantee to the vRAN sufficient resources to respect its real-time constraints as offered with YinYangRAN~\cite{10621380}.
Sharing single-function HAs between vRAN instances is feasible as well
by relying on an AAL capability which allows to balance workloads with the same profile
between several HAs and software libraries under the control of a RAN Intelligent Controller (RIC)
in order to mitigate excess delays.
Ultimately CloudRIC~\cite{10.1145/3636534.3649381} demonstrates that with a computing resource aware MAC scheduling policy, an AAL broker and a real-time RIC
an industry-grade real-time behavior can be achieved for several vRAN instances sharing a HA and a pool of GPP cores.

{\bf Perspectives}
Most of these solutions
may be integrated on top of the OpenAirInterface stack.
Indeed, they usually either rely on elements of the standard O-RAN architecture to control
vRAN instances to share computing resources~\cite{10621380, 10.1145/3636534.3649381}
or implement special scheduling policies on top of the Linux kernel
and CPU states features~\cite{10.1145/3452296.3472894, 10.5555/3767955.3768020}.
Implementing these solutions on top of the framework our work provides
would allow to extensively test, benchmark or even improve them.

\section{Conclusion}

\label{sec:conclusion}

To conclude with our work
on scaling up a vRAN deployment
on top of chosen computer architectures,
we wish to summarize the most important takeaways.

{\bf Achievements}
We demonstrated our ability to operate
a large scale vRAN deployment
with the open-source software stack
OpenAirInterface.
The stack gets a true
advantage from targeted hardware acceleration
compared to the traditional
general-purpose processor based approach.
We achieve a deployment at a scale
unprecedented with this stack.
The deployment achieves
a high-level of resource usage
on large and powerful servers.
Then, the efficiency in term of
computing resource and energy
usage we achieved is also
unprecedented with this stack.

{\bf Challenges for the vRAN}
There are still true challenges to implement
vRANs in operational deployments.
Our deployment still relies on computing resources
dedicated to specific tasks while the few resources
we managed to share experience
overheads due to sharing that threaten
the real-time compliance of vRAN instances. 
Meanwhile, the literature
offers plenty of solutions for vRANs to share compute.
But a framework was maybe missing
to implement and test these solutions
in large-scale end-to-end deployments.
We hope this work could serve as a framework
for researchers and developers
to bring these solutions into operational deployments.

{\bf Computer architectures behaviors}
Finally, our work revealed behaviors of the computer architectures
that network implementers have to adapt to.
The effect of the internal partitioning
in core complexes of the AMD EPYC processors is quite new to us and to
other developers of the OpenAirInterface stack we interacted with.
It adds a grain of salt when thinking the way we use the computing resources,
especially for vRANs since they cannot seamlessly cross core complex
borders to use neighboring resources.
We still need to better understand this behavior and how to work with it.
We also observed that different architectures relying on hardware accelerators
behave differently when they are shared by many vRAN instances.

\section*{Acknowledgments}

This research is funded by the European Union's Horizon Europe research and innovation program
under grant agreement No. 101092598 (COREnext).

\bibliographystyle{plain}
\bibliography{references}


\end{document}